# Hot electroweak matter

K. Kajantie[a]

[a]Department of Physics
P.O.Box 9, FIN-00014 University of Helsinki, Finland
kajantie@phcu.helsinki.fi

This talk summarises recent results on lattice Monte Carlo studies of the finite $T$ electroweak phase transition. Particular attention is given to the 3d effective theory approach, replacing the full 4d theory by a three dimensional effective theory of the modes constant in imaginary time.

## 1. THE PROBLEM

The aim is to solve the following well-defined physical problem:

Given the finite temperature minimal standard model (MSM, electroweak theory, EW theory), parametrised by the – so far unknown – Higgs mass $m_H$, does there exist a symmetry breaking/restoring phase transition at some temperature $T_c$ and if so, what are firstly the properties of the transition itself:
1. order, first or second;
2. numerical value of $T_c$;
3. order parameter jump $v(T_c)$;
3. latent heat $L$, interface tension $\sigma$ and various correlation lengths $\xi_i(T_c)$;
and, secondly, what are the properties of the hot fluid away from $T_c$: equation of state, highest superheating and lowest supercooling temperatures $T_+$ and $T_-$, correlation lengths and the "magnetisation curve" $v(T)$?

A definite quantitative answer to these questions is obviously necessary for any definite quantitative computation of the net baryon number $\Delta B$ generated in the cosmological EW transition [1]. The methods developed for the hot minimal EW theory should also be easily extensible to any hot beyond the standard model theory, if any such is found necessary for generating $\Delta B$.

To focus on the essentials one further simplifies the MSM to a minimal minimal standard model by neglecting the U(1) part of the gauge symmetry group, i.e., by taking $g' = 0$, $\theta_W = 0$ and by neglecting fermions entirely. The theory considered then becomes an SU(2) Higgs theory with a gauge field $A_\mu^a$ and a Higgs field $\phi = (\phi^+, \phi^0)$ in the fundamental representation of the gauge group.

In view of the great complications caused by the quarks for the study of the finite $T$ QCD phase transition and the failure to present any method of putting chiral fermions on the lattice, one asks for the justification of the neglect of fermions. In fact, without this justification it would be impossible to compute the properties of the finite $T$ EW transition: one needs nonperturbative methods but cannot regulate the theory with chiral fermions nonperturbatively. The answer, in formal terms, is that for the EW theory dimensional reduction [2–5], replacement of the full 4d theory by an effective 3d theory, works. The fermion thermal masses $(2n+1)\pi T$ are of the same order as the masses of the nonstatic boson fields $2n\pi T$ and both can either be integrated out, leaving the 3d static massless modes (as in EW theory, also for $T = T_c$ and below) or both have to be included (as in QCD at $T_c$). Fermions only affect the couplings of the effective 3d theory in a simple way, they do not appear as dynamical fields.

In contrast to the extensive work devoted to finite $T$ QCD there has been little work on hot EW theory. The basic formalism was set up and applied in [6] and the phase structure of the lattice action was studied in [7]. The problem was resurrected in [8] and further studied in more extensive and still ongoing numerical 4d simulations by the DESY group [9–11]. The use of 3d simu-lations was initiated in [12] and reviewed in [13], the theory was developed in [14,15] and further simulations were carried out in [16]. A 3d effective theory containing only the Higgs field was formulated and simulated in [17,18].

The motivation and accuracy of the $4d \to 3d$ reduction is clarified by the following discussion.

## 2. 3d EFFECTIVE THEORIES

In physics it is natural to focus on the essentials, to keep only the essential degrees of freedom and to integrate over the irrelevant ones. In the context of finite $T$ SU(2) Higgs theory one can develop the following hierarchy of theories [19].

The 4d theory is defined by an action

$$S[A^a_\mu(\tau,\mathbf{x}),\phi_k(\tau,\mathbf{x})], \tag{1}$$

where the field configurations are periodic over the imaginary time interval $0<\tau<\hbar/T$. Here the modes varying as a function of $\tau$ or, in Fourier analysis, the modes with index $n=\pm1,\pm2,..$ have a mass $2n\pi T$. If now $2\pi T$ is large, relative to the relevant mass scales (inverse correlation lengths at $T$), these nonstatic modes can be integrated over and a new effective theory can be defined by the action $S_{\rm eff}[A^a_i(\mathbf{x}),A^a_0(\mathbf{x}),\phi_k(\mathbf{x})]$.This is an SU(2) + adjoint Higgs + fundamental Higgs theory with coefficients determined by perturbation theory. Further one observes that in some cases the Debye mass $m_D=\sqrt{5/6}gT$ is large. Then one can also integrate over $A^a_0$ and obtain a still simpler effective theory

$$S_{\rm eff}[A^a_i(\mathbf{x}),\phi_k(\mathbf{x})], \tag{2}$$

again with determined coefficients. Finally, it may appear that either $m_H(T)$ or $m_W(T)$ is large. In the former case an effective action $S_{\rm eff}[A^a_i(\mathbf{x})]$ would be obtained, in the latter an effective action

$$S_{\rm eff}[\phi_k(\mathbf{x})], \tag{3}$$

In more detail, the Lagrangian of the full 4d theory in eq.(1) is given by

$$L = \frac{1}{4}F^a_{\mu\nu}F^a_{\mu\nu}+(D_\mu\phi)^\dagger(D_\mu\phi) -\frac{1}{2}m^2\phi^\dagger\phi+\lambda(\phi^\dagger\phi)^2. \tag{4}$$

Integrating over the non-static modes to 1-loop accuracy in the $\overline{\rm MS}$ scheme, which is just a mechanical well controllable computation in perturbation theory, one obtains the following effective action:

$$S_{\rm eff}[A^a_i,A^a_0,\phi_k]=\int d^3x\Big\{\frac{1}{4}F^a_{ij}F^a_{ij}+ \frac{1}{2}(D_iA_0)^a(D_iA_0)^a+(D_i\phi)^\dagger(D_i\phi)+ +\frac{1}{2}m_D^2A^a_0A^a_0+\frac{1}{4}\lambda_A(A^a_0A^a_0)^2+ +m_3^2\phi^\dagger\phi+\lambda_3(\phi^\dagger\phi)^2+h_3A^a_0A^a_0\phi^\dagger\phi\Big\}. \tag{5}$$

The couplings here are given in terms of the 4d couplings by

$$g_3^2=g^2(\mu_T)T, \tag{6}$$

$$\lambda_3=T\left[\lambda(\mu_T)+\frac{1}{16\pi^2}\frac{3}{8}g^4(\mu_T)\right], \tag{7}$$

$$h_3=\frac{1}{4}g_3^2\left[1+\frac{1}{16\pi^2}\left(12\lambda(\mu_T)+\frac{47}{6}g^2(\mu_T)\right)\right], \tag{8}$$

$$\lambda_A=\frac{17g^4(\mu_T)T}{48\pi^2}, \tag{9}$$

$$m_D^2=\frac{5}{6}g^2(\mu_T)T^2, \tag{10}$$

$$m_3^2=\left[\frac{3}{16}g^2(\mu_T)+\frac{1}{2}\lambda(\mu_T)\right]T^2-\frac{1}{2}m_H^2, \tag{11}$$

where

$$\mu_T=4\pi Te^{-\gamma}\approx 7T \tag{12}$$

and the 4d couplings are run to this scale by the standard $\beta$ functions. The appearance of this scale follows from the fact that using the $\overline{\rm MS}$ scheme always introduces an undetermined normalisation scale $\mu$ and terms involving $\log\mu$. The scale choice (12) simply minimises these logarithmic terms. Note the large factor 7.

An essential property of 3d theories is their superrenormalisability: the couplings do not run, only the masses run due to linearly divergent 1-loop and logarithmically divergent 2-loop diagrams. The result for the Higgs mass is

$$m_3^2(\mu_3) = \frac{1}{16\pi^2} f_{2m} \log\frac{\Lambda_m}{\mu_3}, \tag{13}$$

where $\Lambda_m$ is a renormalisation group invariant scale and

$$\begin{aligned} f_{2m} &= \frac{81}{16} g_3^4 + 9\lambda_3 g_3^2 - 12\lambda_3^2 \\ &= T^2\left[1-\left(\frac{m_H}{3m_W}\right)^2\right]\left[3\left(\frac{m_H}{3m_W}\right)^2+1\right], \end{aligned} \tag{14}$$

where on the second line one has inserted $g_3^2 = g^2T$, $g = 2/3$. This is an exact result with no further perturbative correction. Note that for $m_H = 3m_W$ even the Higgs mass does not run.

As discussed in [14,15,19], in the limit of large $m_D$ the triplet scalar field $A_0$ can be integrated out from the theory by simply removing it from the action (5) and making simple and small changes in $g_3^2, \lambda_3, m_3^2$. Since the essential dynamics lies in the $A_i, \phi$ sector, this approximation is the one to make for the study of the physical EW case. It simplifies the interpretation of lattice Monte Carlo simulations considerably by containing one mass scale less.

Finally, the action obtained by integrating the theory defined by eq.(5) over the gauge field $A_i^a$ is [17]

$$\begin{aligned} &S[\phi_k(\mathbf{x})] = \\ &\int d^3x \Big\{ \partial_i\phi^\dagger\partial_i\phi + m_3^2\phi^\dagger\phi + \lambda_3(\phi^\dagger\phi)^2 \\ &-\frac{1}{4\pi}g_3^2\left[2\left(\frac{1}{2}\phi^\dagger\phi\right)^{3/2} + \left(m_D^2+\frac{1}{2}\phi^\dagger\phi\right)^{3/2}\right]\Big\}. \end{aligned} \tag{15}$$

## 3. VALIDITY OF THE 3D APPROXIMATION

In physical terms, the 3d approximation is valid when $T$ is "large". Here "large" must mean relative to all relevant mass scales of the problem. In the finite $T$ context these scales are screening lengths and inverse correlation lengths. Without any extra scales from symmetry breaking all mass scales are proportional to powers of coupling constants times $T$. This leads to the first condition

$$g^2(\mu_T), \lambda(\mu_T) \ll 1. \tag{16}$$

Equivalently, this is the condition that the 1-loop perturbative computation leading from the full theory in eq.(4) to the effective theory (5) be valid. It is the only condition for QCD and for EW theory in the symmetric high $T$ phase.

For EW theory symmetry breaking in the low $T$ phase brings with it further mass scales and, to use the approximation near $T_c$, one must also demand that

$$T_c \gg 1/\xi_W(T_c),\ 1/\xi_H(T_c). \tag{17}$$

When translated to $m_H$ these two conditions imply that the theory in eq.(5) is valid when

$$30\,\mathrm{GeV} \lesssim m_H \lesssim 240\,\mathrm{GeV}. \tag{18}$$

The vaguely defined upper limit expresses the usual fact that for large Higgs masses EW theory becomes strongly coupled, the lower comes from eq.(17) – $T_c$ goes down with small $m_H$ while simultaneously the other scales increase since $v(T)$ approaches $v(0) = 246$ GeV.

For QCD there is now ample evidence of the fact that the 3d approximation works down to $T$ rather close to $T_c$, perhaps for $T \gtrsim 1.5T_c$. In fact, there is even quantitative evidence [20] that this is related to the large thermal scale $\mu_T$ in eq.(12): at $T_c$ the QCD coupling is then smaller than one might expect. However, at $T_c$ quarks become essential for the transition dynamics and the 4d→3d reduction does not work. The dominant configurations must then vary with $\tau$ and their shapes have been studied in [21].

Reduction to 3d thus works when couplings are small at the scale $\approx 7T$, but another crucial aspect is that it also works when 4d finite $T$ perturbation theory does NOT work. The criterion for $T \neq 0$ perturbation theory to work is namely that

$$\frac{g^2T}{2\pi Q} \ll 1, \tag{19}$$

where $Q$ is some relevant mass scale, $Q = k$ or $Q = m_W(T) = gv(T)/2$. The derivation of the effective 3d theory does not care about this condition. Of course, if we want the study the 3d theory perturbatively, we are back to eq.(19), but we can as well study the effective theory nonperturbatively with lattice Monte Carlo, it contains all the infrared problems of finite $T$ 4d theory. The difficulty has thus been cornered.

## 4. WHY PERTURBATION THEORY IS NOT SUFFICIENT

There is one more point one should clearly appreciate: perturbation theory may very well work in the broken low $T$ phase (it works extremely well for $T=0$ with $v(T=0)=246$ GeV so why not as long as $v(T)$ remains clearly nonzero?) but it cannot work in the high $T$ symmetric phase with $v(T)=0$. It thus cannot work for anything requiring a comparison of the symmetric and broken phases. For example, $T_c$ is determined from $p_{\rm symm}(T_c)=p_{\rm broken}(T_c)$. This also holds for $L,\sigma$ and $\xi_{\rm symm}(T)$. Of course, low order perturbation theory will give qualitative estimates.

In some more detail, in successive orders of the loop expansion the terms of the effective potential go like

$$\frac{1}{4}\lambda_3\phi^4,\quad \frac{m_W^3}{2\pi},\quad \frac{g_3^2m_W^2}{(2\pi)^2},\quad \frac{g_3^4m_W}{(2\pi)^3},\quad \frac{g_3^6}{(2\pi)^4},\quad \ldots, \tag{20}$$

where $m_W=g_3\phi/2$. The potential up to 2 loops is known [14,22–26] and includes terms on the first line. Perturbation theory is valid when the expansion parameter (19) is small:

$$\frac{g_3^2}{2\pi m_W}\ll 1,\quad \phi\gg\frac{1}{\pi}gT, \tag{21}$$

and near $\phi=0$ perturbation theory always has an error

$$\Delta V(0)\sim V_{\rm 2loop}(\frac{gT}{\pi})\sim V_{\rm 4loop}\sim\frac{g^6T^4}{(2\pi)^4}. \tag{22}$$

This is what one has to improve with lattice Monte Carlo computations.

## 5. THE LATTICE ACTION

The lattice action corresponding to the 4d continuum theory in eq.(4) and the 3d effective continuum theory in eq. (5) (the five terms [12] containing the $A_0^a$ field are not written explicitly, for brevity) has the following standard form:

$$S=\beta_G\sum_x\sum_{i<j}(1-\frac{1}{2}{\rm Tr}P_{ij})+\sum_x\Big[\frac{1}{2}{\rm Tr}\,\Phi^\dagger(x)\Phi(x)+\beta_R\big(\frac{1}{2}{\rm Tr}\,\Phi^\dagger(x)\Phi(x)-1\big)^2-\beta_H\sum_i\frac{1}{2}{\rm Tr}\,\Phi^\dagger(x)U_i(x)\Phi(x+i)\Big]. \tag{23}$$

Here $U_i(x)$ and $P_{ij}$ are the standard link and plaquette variables and the scalar field is also represented by a 2×2 matrix $\Phi=R_LV$, $R_L$ = radial mode, $V$ = element of SU(2). In the notation of [10], $\beta_G\to\beta$, $\beta_H\to 2\kappa$ and $\beta_R\to\lambda$. The only difference between the 4d and 3d cases at this stage is in the range of $i,j$: 1 to 4 or 1 to 3.

The system is now simulated for various values of coupling constants, expectation values of various gauge invariant operators are evaluated and signals for a phase transition are searched for. The crucial question then is how the values of $\beta_H,\beta_R,\beta_G$ are converted to physical values of $m_W,m_H,T$ and what operators should be used to measure $L,\sigma,v(T)$ and correlation lengths. It is here that the difference between the 4d and 3d cases is large.

The analysis starts from the tree relations between the lattice and continuum coupling constants. These are

$$\beta_G = \frac{4}{g^2}\quad \text{(4d case)}, \tag{24}$$

$$= \frac{4}{g_3^2}\frac{1}{a}\quad \text{(3d case)}, \tag{25}$$

and, for a potential of the form $\frac{1}{2}m^2\phi^2+\frac{1}{4}\lambda\phi^4$,

$$\beta_R = \frac{1}{4}\lambda\beta_H^2\quad \text{(4d case)}, \tag{26}$$

$$= \frac{1}{4}\lambda_3a\beta_H^2\ \text{(3d case)}, \tag{27}$$

and ($d$ = 3,4)

$$m^2=\frac{2(1-2\beta_R-d\beta_H)}{\beta_Ha^2}. \tag{28}$$

As a concrete example of how the signal for the phase transition appears, Fig. 1 shows the distribution in $\langle\frac{1}{2}{\rm Tr}\,\Phi^\dagger\Phi\rangle$ in the 3d case on an $N^3=24^3$ lattice for fixed $\beta_G=12,\beta_R=0.00126$ (corresponding to $m_H=80$ GeV) and for varying $\beta_H$. For illustration, the values of $T$ corresponding to these $\beta_H$ would in the $a \to 0$ limit – using the constant physics curve given below – be those in the caption. One sees how the single peak corresponding to the broken phase develops into a two-peak structure indicating a transition. The situation even closer to $T_c$ and on both sides of it is shown in Fig. 2. Both $\beta_{Hc}$ and hence $T_c$ are very accurately determined – depending on how accurately the constant physics curve is known.

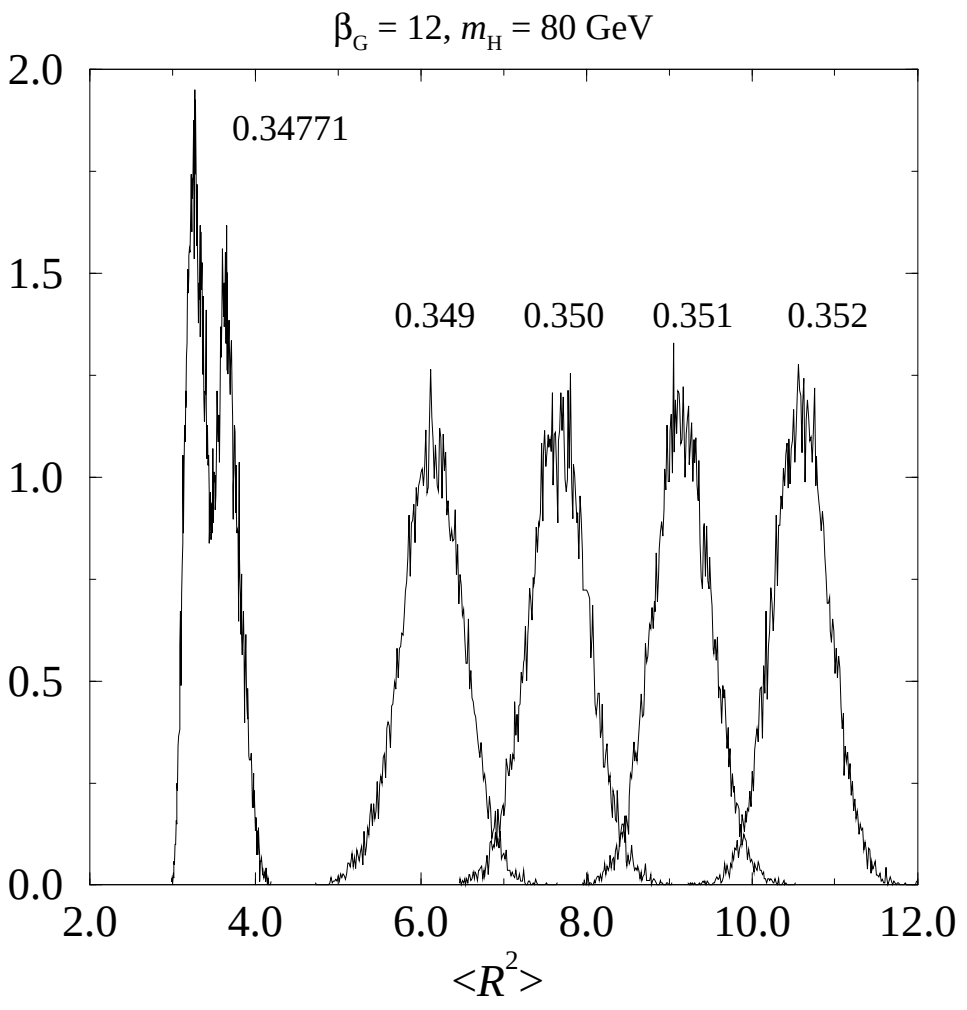

Figure 1. Distributions of $\langle R^2 \rangle = \langle \phi^\dagger \phi \rangle$ for $m_H = 80$ GeV, $N^3 = 24^3$, $\beta_G = 12$, for varying $\beta_H$ below and at the phase transition. Using the constant physics curve (37) the values of $\beta_H$ would for $a \to 0$ correspond to (from the left) $T = 171.3$, 147.6, 134.8, 124.9, 116.9 GeV.

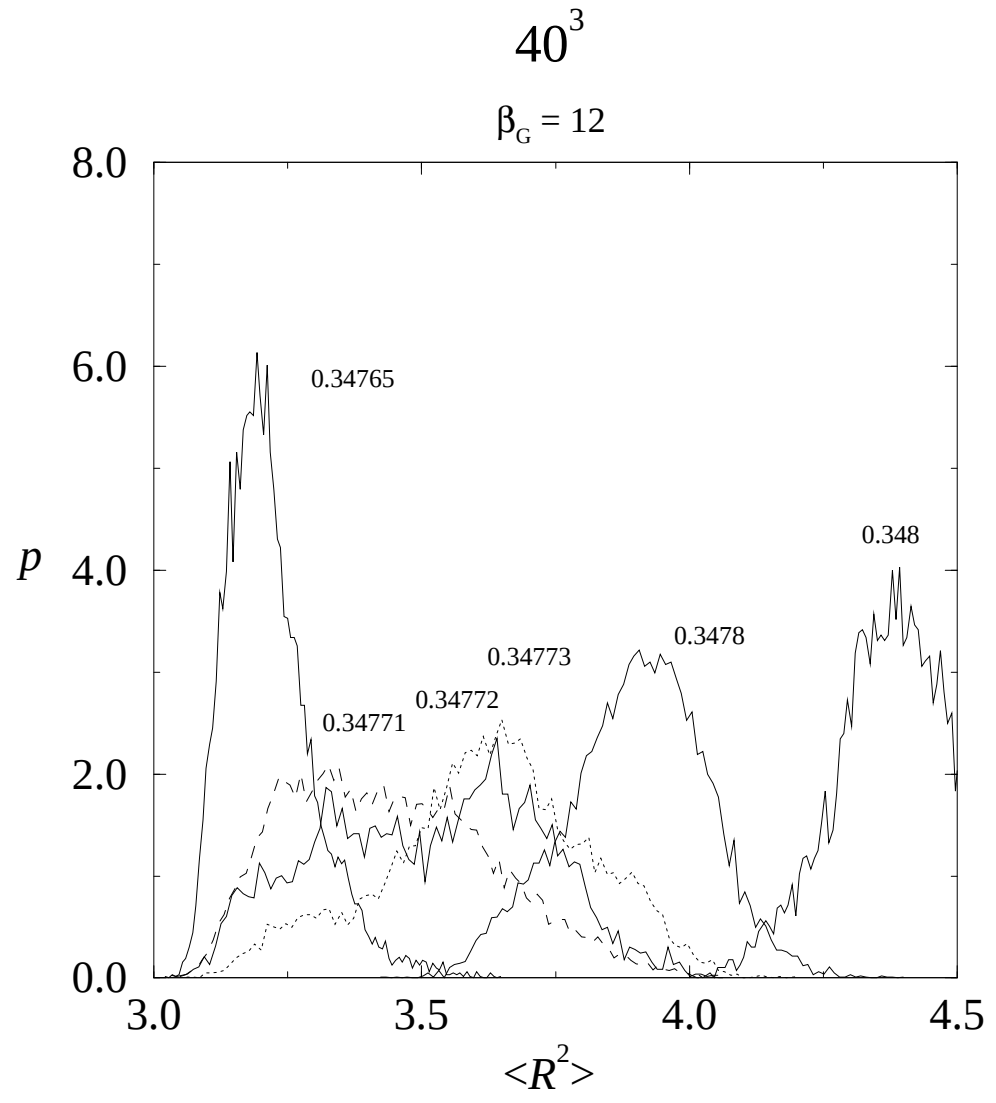

Figure 2. Distributions of $\langle R^2 \rangle = \langle \phi^\dagger \phi \rangle$ for $m_H = 80$ GeV, $N^3 = 24^3$, $\beta_G = 12$, for varying $\beta_H$ very close to the phase transition. Using the constant physics curve (37) the values of $\beta_H$ would for $a \to 0$ correspond to (from the left) $T = 172.7$, 171.3, 171.0, 170.8, 169.3, 164.9 GeV.

### 5.1. 4d case

In the 4d case the lattice parameters are converted into physical numbers by the following procedure [9–11]. First one chooses an $N_t N_x^2 N_z$ lattice with $N_t \ll N_x, N_z$ and $N_z \gg N_x$ (for interface tension measurements). The values used were $N_t = 2, 3$, the spatial sizes varied from $16^2{\cdot}24$ to $96^2{\cdot}192$. Then simply

$$\frac{1}{T} = N_t a \tag{29}$$

and the problem has been converted to determining the lattice spacing $a$. For the bare couplings one chooses

$$\beta_G = 8 \tag{30}$$

and

$$\beta_R = 0.0001 \quad \text{and} \quad \beta_R = 0.0005. \tag{31}$$

It is here that the newest simulations differ from the early ones [6,7], which used smaller $\beta_G$ (closer to the $\beta_{Gc} \approx 2.3$ of pure finite $T$ SU(2) gauge theory) and larger $\beta_R$ ($\beta_R = 0.5$, say). The reason for this choice [8] is that these are in the range of the values following from the tree relations in eqs.(24-27) using the standard model values $g^2 \approx 0.5, \lambda = g^2 m_H^2/(8 m_W^2)$ (with $m_H = 18$, 50 GeV). Renormalisation effects are observed to be small.

As an example of the distributions obtained, Figs. 3 and 4 show $\langle \phi^\dagger \phi \rangle$ for $m_H = 18$ GeV in two parts, showing clearly the well separated symmetric and broken phase peaks at $T_c$. A reweighted unified distribution is shown in [10,11].

In the $V \to \infty$ limit, for $N_t = 2$, one finds

$$\beta_{Hc} \quad = \quad 0.128290(1) \quad (\beta_R = 0.0001)$$

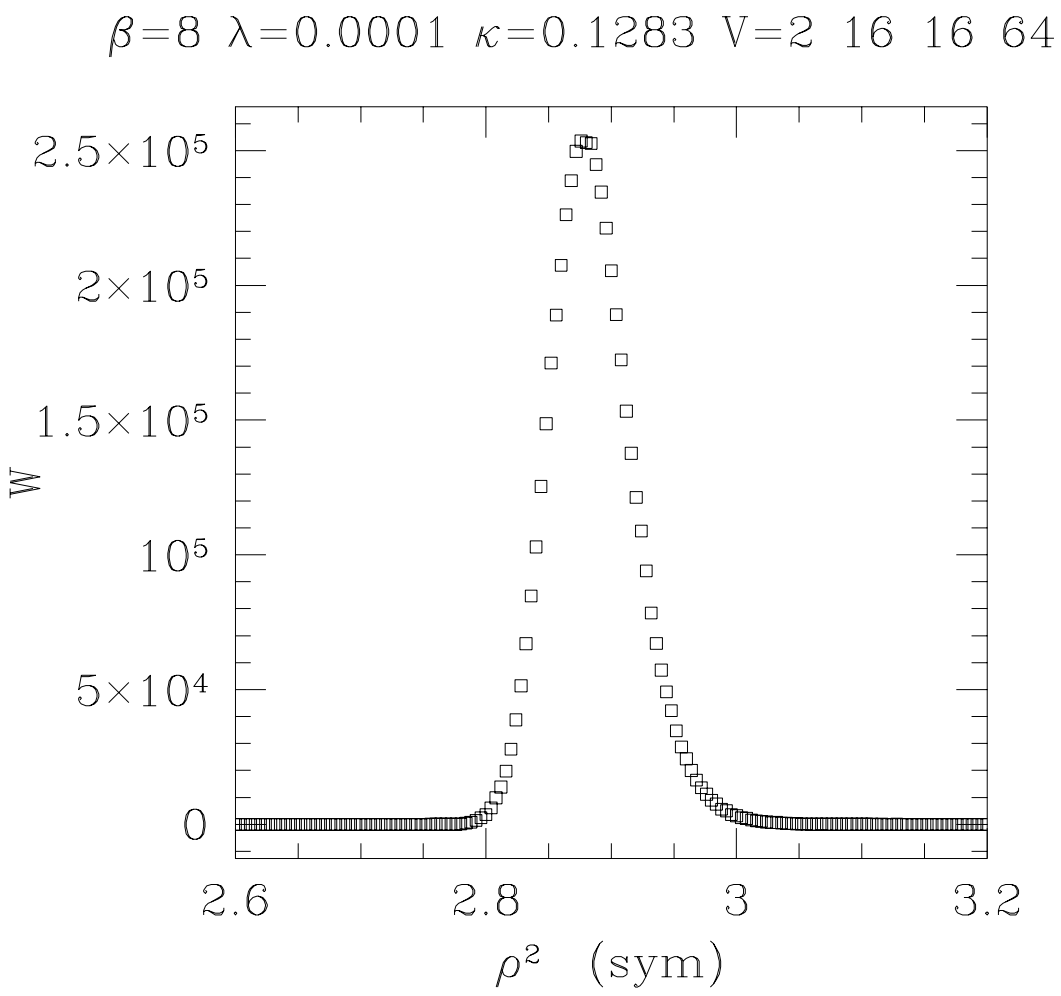

Figure 3. Distributions of $\rho^2 = \langle\phi^\dagger\phi\rangle$ in the 4d theory for $m_H = 18$ GeV in the symmetric phase. Data from the DESY group.

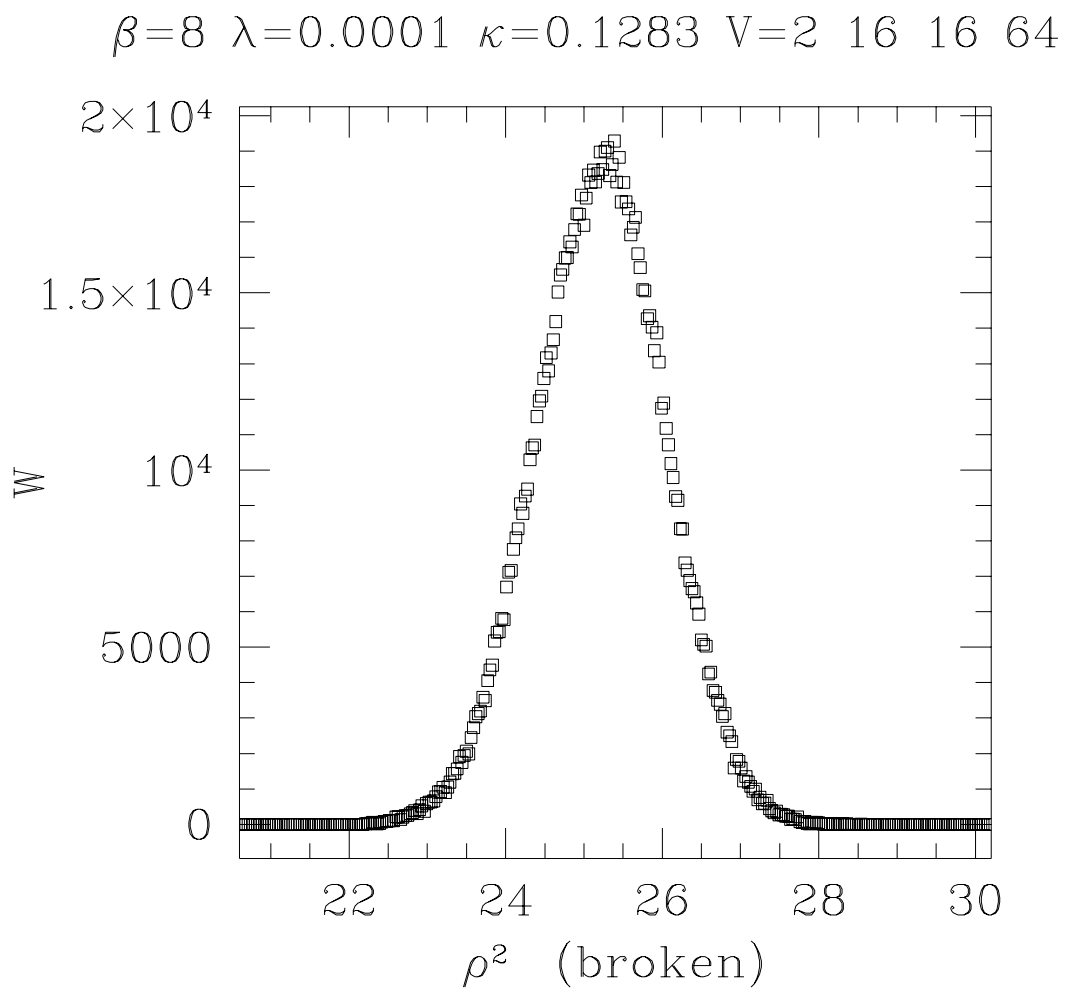

Figure 4. As Fig.3 but in the broken phase. Note the change of scale from Fig. 3.

$$= \quad 0.128625(1) \quad (\beta_R = 0.0005). \tag{32}$$

As a second step, one goes to a symmetric $N_t^4$ lattice (or to lattices with $N_t > N_s$) and determines both $am_W$ and $am_H$ by using suitable correlation functions. Combining with determination of $\beta_{Hc}$ and $1/T = N_t a$ one then has

$$T_c/m_W, \qquad m_H/m_W \tag{33}$$

for finite $a$. As a final step, one then runs to $a \to 0$ using known 1-loop beta functions:

$$-a\frac{dg}{da} = \beta_g(g)$$
$$-a\frac{d\lambda}{da} = \beta_\lambda(g^2,\lambda). \tag{34}$$

Good scaling means that $1/T_c = N_t a$ computed for various $N_t$ is the same, i.e., the change in $a$ compensates for that in $N_t$. Computing the change in $a$ caused by $N_t = 2 \to 3$, then from eq.(34) the change in the bare parameters caused by this change in $a$ and finally performing simulations at these new values of bare parameters, one finds [10] that scaling between $N_t = 2$ and 3 is actually surprisingly well satisfied, in constrast to QCD. This again indicates that the modes constant in imaginary time are the dominant ones.

### 5.2. 3d case

In the 3d case one chooses an $N^3$ lattice, $N = 8, ..., 48$ have been used. Then the lattice spacing is immediately fixed by eqs. (6) and (25):

$$a = \frac{4}{g^2\beta_G}\frac{1}{T}. \tag{35}$$

This shows explicitly how the continuum limit is given by $\beta_G \to \infty$. Similarly, eqs. (7) and (27) fix $\beta_R$:

$$\beta_R = \frac{\lambda_3}{g_3^2}\frac{\beta_H^2}{\beta_G} = \frac{1}{4}\lambda\beta_H^2 Ta. \tag{36}$$

Due to superrenormalisability, there are no radiative corrections to these relations. The main relation, the constant physics curve, relates $\beta_H$ to the other couplings and the physical quantities, $T$, $m_W$, $m_H$ and follows after introducing radiative corrections [27] to the tree level mass relation (28). For a theory with the $A_i^a, \phi, A_0^a$ fields the constant physics curve is (if $A_0^a$ is integrated over, see [15])

$$\frac{m_H^2}{4T^2} = \left(\frac{g^2\beta_G}{4}\right)^2\Bigg[3 - \frac{1}{\beta_H} + \frac{m_H^2}{4m_W^2}\frac{\beta_H}{\beta_G}$$
$$-\frac{9}{8\pi\beta_G}\left(1+\frac{m_H^2}{3m_W^2}\right)\Sigma -$$

$$-\frac{1}{2}\left(\frac{9}{4\pi\beta_G}\right)^2\left\{\left(1+\frac{2m_H^2}{9m_W^2}-\frac{m_H^4}{27m_W^4}\right)\log\frac{g^2\beta_G}{2}\right.$$
$$\left.+\eta+\frac{2m_H^2}{9m_W^2}\bar{\eta}-\frac{m_H^4}{27m_W^4}\tilde{\eta}\right\}\Bigg]$$
$$+\frac{g^2}{2}\left[\frac{3}{16}+\frac{m_H^2}{16m_W^2}+\frac{g^2}{16\pi^2}\left(\frac{149}{96}+\frac{3m_H^2}{32m_W^2}\right)\right]$$
$$+\mathcal{O}(Ta). \qquad (37)$$

In this equation, the first line is essentially the tree relation (28), the second line is the linearly divergent 1-loop mass counter term, $\Sigma = 3.175911$ being a constant related to its regulation in the lattice scheme, the third line is the leading 2-loop logarithmic counter term, the fourth line contains the constants $\eta \approx 2.2$, $\bar{\eta} = 1.01$, $\tilde{\eta} = 0.44$ appearing when one relates the 2-loop mass renormalisation in the continuum $\overline{\rm MS}$ and the lattice regularisation schemes, the fifth line contains the constant $\gamma$ (with a 2-loop correction) of the thermal $\gamma T^2$ mass term and finally the last line emphasises the fact that when applying this formula to lattice data one should perform an extrapolation to the $a \to 0$ ($\beta_G \to \infty$) limit.

The constants in (37) are analogous to the numbers $\Lambda_{\overline{\rm MS}}/\Lambda_{\rm Lattice}$ [28,29] in QCD with the difference that in QCD one has to carry out a 4d 1-loop and here a 3d 2-loop calculation. Thus the present situation is more complicated and only the numbers $\bar{\eta}$ and $\tilde{\eta}$ have been computed analytically but not $\eta$ yet. However, the value $\eta \approx 2.20$ can be obtained by calibrating the constant physics curve so that a lattice measurement of $\langle\phi^\dagger\phi\rangle$ gives the same result as a continuum 2-loop computation thereof. The latter contains explicitly $T$ (or, actually, the couplings of the 3d theory) and a comparison permits one to calibrate the relation between $T$ and the lattice quantities. Sample data was shown in Fig.1. When all these three constants are known, the constant physics curve in eq.(37) is exact in the limit $a \to 0$: there are no further perturbative corrections to it.

## 6. RESULTS

At the time of the conference the situation is developing rather rapidly and present results will soon be replaced by more accurate ones and new regions of parameter space will be explored. Thus only a short summary with rounded numbers will be given here. Nevertheless, it is interesting to contrast the level of accuracy one is aiming at in EW theory and in QCD (see, in particular, Fig. 2). Due to the weak coupling nature of the theory one is discussing values of physical quantities to three or even four significant digits, which is unheard of in finite $T$ QCD.

### 6.1. Simulations in the broken phase

One expects both perturbation theory to work well in the broken phase (since the large value of $\phi$ will shield all infrared singularities at $\mathbf{k} \to 0$) and simulations to be rather straightforward (small autocorrelation times). This is what happens; the data deep in broken phase in Fig. 1 could be produced with relatively little computational effort. Using data of this type for $\langle\phi^\dagger\phi\rangle$ and requiring it to be the same on the lattice and in perturbation theory can then be used to calibrate the lattice $\leftrightarrow$ physics relation in the 3d theory. Other similar condensates could be used, but no 2-loop perturbative computations exist for them as yet.

### 6.2. Order of transition, $T_c$ and $v(T_c)$

The properties of the transition have been studied for Higgs masses in the range of 18 to 80 GeV. The transition is strongly first order in the lower end of this mass range but gets rapidly weaker when $m_H$ increases. The strength for $m_H = 18$ GeV is illustrated by the well separated two peaks in Figs. 3 and 4 [9–11]. For $m_H = 35$ GeV a proof of the 1st order nature based on the volume dependence of the dispersion of the single link operator is shown in Fig. 5 [12]. For $m_H = 80$ GeV rigorous tests of this type have not yet proven the order of the transition.

The above result is in qualitative agreement with perturbation theory. Quantitatively the transition seems to be stronger than predicted by perturbation theory. Various estimates of higher loop corrections have also predicted the disappearance of the transition when $m_H$ becomes larger than 70 or 80 GeV.

Some quantitative numbers are given in Table 1. There is one significant conclusion one can draw on the basis of the 35 GeV data in this Table: the transition in an effective theory with only the Higgs field is clearly weaker ($v(T_c)/T_c$ is smaller) than in the theory with also $A_i^a$. This proves that $A_i^a$ plays a significant role in the dynamics of the transition [17].

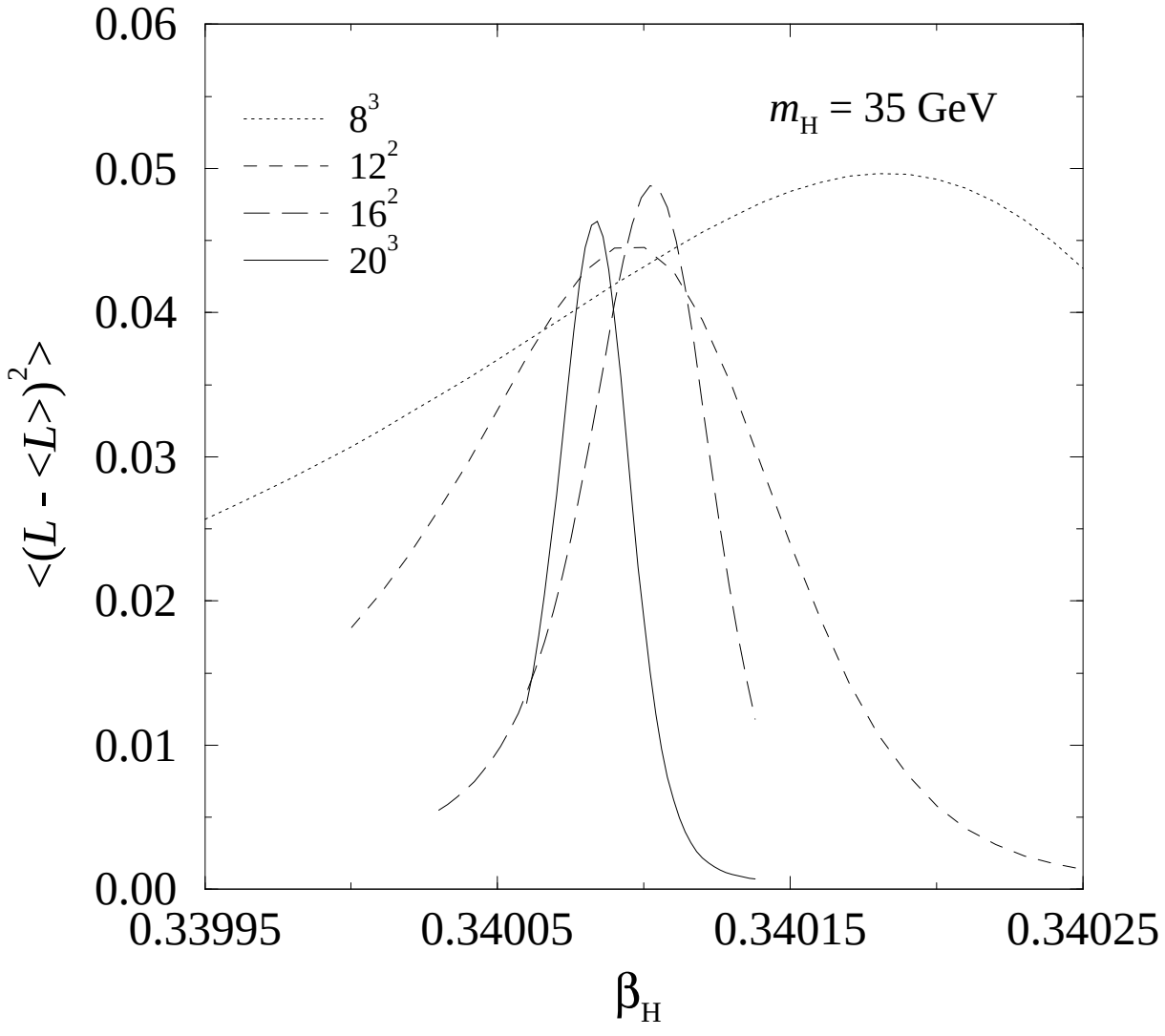

Figure 5. Volume dependence of the dispersion of the link operator $L$ at $T_c$ for $m_H = 35$ GeV [12].

Table 1
Numerical values of $T_c$ and $v(T_c)/T_c$ for various $m_H$ values obtained in the simulations in the first column. Units are GeV.

| simulation | $m_H$ | $T_c$ | $v(T_c)/T_c$ |
|---|---|---|---|
| 4d [9,10] | 18 | 38 | |
| 3d full [12] | 35 | 93 | 1.8 |
| 3d, only $\phi$ [17] | 35 | 115 | 0.7 |
| 4d [9,10] | 49 | 94 | 1.5 |
| 3d full [16] | 80 | 168 | 0.7 |

Secondly, as indicated earlier, fermions are not important for the dynamics of the transition, i.e., they do not affect the nonperturbative simulations but may change their analytic interpretation. The effect of the top quark is most significant: $m_{\rm top} = 175$ GeV changes for $m_H = 80$ GeV the $T_c$ of 167 GeV in Table 1 to 110 GeV, but leaves $v(T_c)/T_c$ almost unchanged.

It may be of interest to consider the dominant energies of microscopic subprocesses in the hot plasma at $T_c$, remembering that the average energy of a boson is about $3T$. For $m_H = 80$ GeV the average initial energy with the above $T_c$ thus is approximately 500 GeV + 500 GeV, a very large number. In the realistic case with top quark this goes somewhat down to 300 GeV + 300 GeV.

### 6.3. Latent heat

The latent heat,

$$L = T_c[p'_{\rm symm}(T_c) - p'_{\rm broken}(T_c)], \quad p_{\rm symm}(T_c) = p_{\rm broken}(T_c), \tag{38}$$

is a physical quantity and can thus be related to expectation values of gauge invariant operators. In the full 4d theory the result [6],[9] contains discontinuities of various terms in the action in eq.(23) multiplied by the derivatives $-a\,d/da$ along lines of constant physics. In the 3d effective theory the result is very simple:

$$\frac{L}{T_c^4} = \frac{m_H^2}{T_c^2}\frac{1}{8}g^2\beta_H\beta_G\Delta\langle\frac{1}{2}{\rm Tr}\,\Phi^\dagger\Phi\rangle. \tag{39}$$

The latent heat thus is directly related to the discontinuity of $\langle\frac{1}{2}{\rm Tr}\,\Phi^\dagger\Phi\rangle$ and can directly be read from the plot.

The following results are obtained:

$$\begin{aligned} \frac{L}{T_c^4} &= 1.6 \quad 4d,\ m_H = 18\,{\rm GeV}, \\ &= 0.26 \quad 3d,\ m_H = 35\,{\rm GeV}, \\ &= 0.12 \quad 4d,\ m_H = 49\,{\rm GeV}, \\ &< 0.03 \quad 3d,\ m_H = 80\,{\rm GeV}. \end{aligned} \tag{40}$$

The errors in the 4d case are 10...20%. The simple way in which the 3d number for 35 GeV is obtained is shown in Fig. 6. For 80 GeV the peaks are not so well separated (Fig. 1) and only an upper limit is obtained.

### 6.4. Interface tension

Precisely at $T_c$ of a first order transition there is a possible two-phase configuration, in which low-$T$ and high-$T$ bulk domains coexist with an interface separating them. The configuration is both thermodynamically ($T = T_c$ the same on both sides) and mechanically ($p = p_c$ the same) stable. The interface tension $\sigma$ is the extra free energy per area it costs to build this interface.

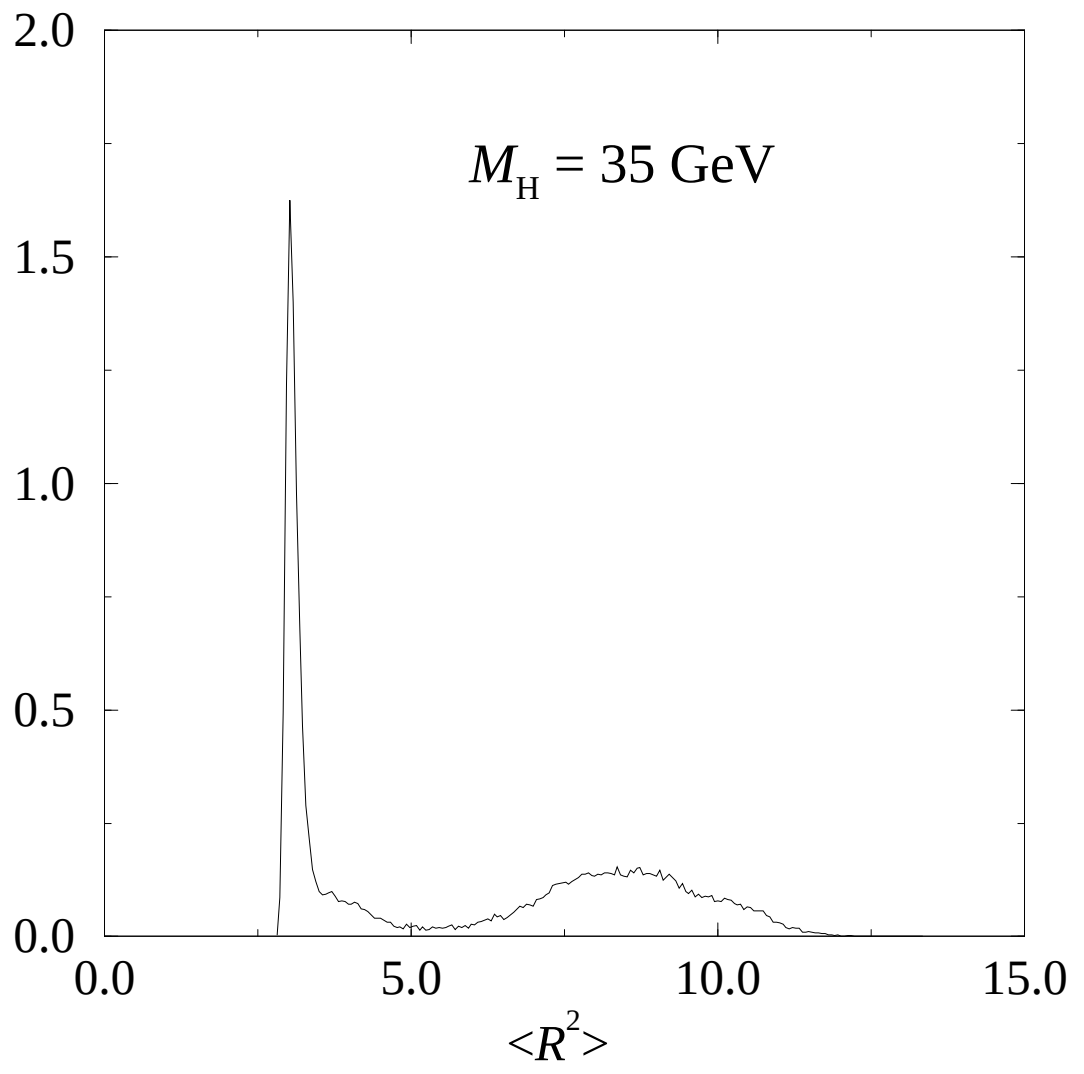

Figure 6. Distribution of $\langle R^2\rangle = \langle\phi^\dagger\phi\rangle$ at $T_c$ (areas under the peaks are equal) for $m_H = 35$ GeV, $\beta_G = 20, \beta_H = 0.34009, N^3 = 20^3$. The averages of the well separated peaks are 3.22 (symmetric) and 8.18 (broken phase). Using $T_c = 93$ GeV in eq.(39) gives the latent heat in (40)

There exist several methods for determining $\sigma$, for example, the two-phase method [30,31] (which involves the creation of the interface by a gradient and then extrapolating to zero gradient) and the histogram method [32] (which determines $\sigma$ by a direct measurement of the extra free energy cost). At present the former two have been applied in the full 4d EW theory [9–11] and the histogram method in the 3d pure Higgs effective theory [17].

Results are given in Table 2. Further theoretical predictions can be found in [24]. One observes, in particular, that the two methods for obtaining $\sigma$ agree with each other very well at 18 GeV. Secondly, the value of $\sigma$ obtained in the effective theory containing only the Higgs field [17] is very small. This again is indicative of the fact that some essential dynamics is lost when $A_i^a$ is integrated over.

Table 2
Numerical values of $\sigma/T_c^3$ for various $m_H$ values obtained in the simulations ((a),(b) refer to two-phase and histogram methods) or theoretical estimates ((th) refers to 2-loop RG improved perturbation theory [14]) in the first column.

| simulation | $m_H$ | $\sigma/T_c^3$ |
|---|---|---|
| 4d(a) [9,10] | 18 | 0.84(16) |
| 4d(b) [9,10] | 18 | 0.83(4) |
| 3d full (th) | 35 | 0.063 |
| 3d, only $\phi$ [17] | 35 | 0.001 |
| 4d [9,10] | 49 | 0.008(2) |
| 3d full (th) | 80 | 0.0034 |

## 7. FUTURE PROSPECTS AND CONCLUSIONS

Although analytic computations cannot make a definite statement, the expectation from them was that for small $m_H$ the transition would be of first order which with increasing $m_H$ or $\lambda$ would become weaker and weaker. Some estimates further imply that at some $m_H$ it would go over to a second order transition and at still larger $m_H$ there would be no transition at all.

The first part of this scenario has been clearly established also by nonperturbative lattice Monte Carlo simulations with quantitative results for the properties of the transition. This result holds for Higgs masses below the present experimental lower limit and is thus only of theoretical interest. For $m_H = 80$ GeV, which is within experimentally acceptable range, there are indications of a first order transition, but the result has not yet been established by conclusive finite size scaling analysis.

On the theoretical side, the role of 3d effective theories and the justification and effectiveness of their use in the study of weakly coupled EW matter in the phase transition region is clearly understood. This, in particular, implies that the notorious complications caused by quarks in the study of the finite $T$ QCD transition do not enter here.

For the future the course of action is clear: extensive numerical simulations in the range $m_H = 60, 70, 80, 90$, etc. GeV should be performed with careful finite size and $a$ scaling analysis. The most economical and simultaneously reliable effective theory is the one containing only the 3d gauge and Higgs fields: $S[A_i^a(\mathbf{x}), \phi(\mathbf{x})]$. Comparisons between different effective theories and the full 4d theory will offer cross checks of the results. On the theoretical side, one should compute using lattice perturbation theory the 2-loop effective potential of SU(2) Higgs theory. This would analytically complete the determination of the constant physics curve of 3d effective theories of hot EW matter.

## REFERENCES

1. M. Shaposhnikov and G. Farrar, Phys. Rev. D50 (1994) 774; Phys. Rev. Lett. 70 (1993) 2833, Erratum-ibid.71 (1993) 210.
2. P. Ginsparg, Nucl. Phys. B170 (1980) 388.
3. T. Appelquist and R. Pisarski, Phys. Rev. D23 (1981) 2305.
4. S. Nadkarni, Phys. Rev. D27 (1983) 917.
5. N. P. Landsman, Nucl. Phys. B322 (1989) 498.
6. P. H. Damgaard and U. Heller, Nucl. Phys. B294 (1987) 253; Nucl. Phys. B304 (1988) 63; Phys. Lett. B171 (1986) 442; Phys. Lett. B164 (1985) 121.
7. H. G. Evertz, J. Jersák and K. Kanaya, Nucl. Phys. B285 (1987) 229.
8. B. Bunk, E.-M. Ilgenfritz, J. Kripfganz and A. Schiller, Phys. Lett. B284 (1992) 371; Nucl. Phys. B403 (1993) 453.
9. F. Csikor, Z. Fodor, J. Hein, K. Jansen, A. Jaster and I. Montvay, Phys. Lett. B334 (1994) 405.
10. Z. Fodor, J. Hein, K. Jansen, A. Jaster and I. Montvay, DESY Preprint DESY-94-159.
11. Z. Fodor, J. Hein, K. Jansen, A. Jaster and I. Montvay, these proceedings.
12. K. Kajantie, K. Rummukainen and M. Shaposhnikov, Nucl. Phys. B407 (1993) 356.
13. K. Kajantie, in Particles and the Universe, pp. 31 - 42, Proceedings of the 17th Johns Hopkins Workshop on Theoretical Physics, Budapest, 1993, Z. Horváth, L. Palla and A. Patkós, editors, World Scientific, Singapore, 1994.
14. K. Farakos, K. Kajantie, K. Rummukainen and M. Shaposhnikov, hep-ph 9404201, Nucl. Phys. B425 (1994) 67.
15. K. Farakos, K. Kajantie, K. Rummukainen and M. Shaposhnikov, CERN Preprint CERN-TH.7220/94.
16. K. Farakos, K. Kajantie, K. Rummukainen and M. Shaposhnikov, hep-ph 9405234, Phys. Lett. B336 (1994) 494.
17. F. Karsch, T. Neuhaus and A. Patkós, Bielefeld Preprint BI-TP 94/27.
18. F. Karsch, T. Neuhaus and A. Patkós, these preceedings.
19. A. Jakovác, K. Kajantie and A. Patkós, hep-ph 9312355, Phys. Rev. D49, (1994) 6810.
20. F. Karsch, E. Laermann and M. Lütgemeier, Bielefeld preprint BI-TP 94/55; M. Lütgemeier, these proceedings.
21. E.M. Ilgenfritz and E.V. Shuryak, Phys. Lett. B325 (1994) 263; E.V. Shuryak and J.J.M. Verbaarschot, Nucl. Phys. B364 (1991) 255.
22. P. Arnold and O. Espinosa, Phys. Rev. D47 (1993) 3546.
23. A. Hebecker, Z. Phys. C60 (1993) 271.
24. Z. Fodor and A. Hebecker, DESY preprint DESY 94-025, Nucl. Phys. B, to be published.
25. M. Laine, Phys. Lett. B335 (1994) 173.
26. M. Laine, hep-ph 9411252, Helsinki preprint HU-TFT-94-46
27. G. Parisi, Statistical Field Theory, Addison Wesley 1988, Ch. 5.
28. A. Hasenfratz and P. Hasenfratz, Phys. Lett. B93 (1980) 165.
29. D. Gross and R. Dashen, Phys. Rev. D23 (1981) 2340.
30. J. Potvin and C. Rebbi, Phys. Rev. Lett. 62 (1989) 3062
31. K. Kajantie, Leo Kärkkäinen and K. Rummukainen, Nucl. Phys. B333, (1990) 100.
32. K. Binder, Phys. Rev. A25 (1982) 1699.